\documentclass[fleqn,usenatbib]{mnras}

\usepackage{newtxtext,newtxmath}

\usepackage[T1]{fontenc}
\usepackage{ae,aecompl}

\usepackage{graphicx}	
\usepackage{amsmath}	
\usepackage{amssymb}	
\usepackage{tablefootnote}

\title[Revisiting the GRB isotropy test]{Revisiting the statistical isotropy of GRB sky distribution}

\author[Uendert Andrade et al.]{
Uendert Andrade,$^{1}$\thanks{E-mail: uendertandrade@on.br}
Carlos A. P. Bengaly,$^{2}$\thanks{E-mail: carlosap87@gmail.com}
Jailson S. Alcaniz,$^{1,3}$\thanks{E-mail: alcaniz@on.br} \and
Salvatore Capozziello$^{4,5,6,7}$\thanks{E-mail: capozziello@na.infn.it}
\\
\\
$^{1}$Observat\'orio Nacional, Rio de Janeiro, RJ 20921-400, Brasil\\
$^{2}$Department of Physics \& Astronomy, University of the Western Cape, 7535 Cape Town, South Africa,\\
$^{3}$Departamento de F\'isica, Universidade Federal do Rio Grande do Norte, Natal, RN 59072-970, Brasil,\\
$^{4}$Dipartimento di Fisica ``E. Pancini``,  Universit\`a di Napoli ``Federico II'',\\
Complesso Universitario di Monte Sant' Angelo, Edificio G, Via Cinthia, I-80126, Napoli, Italy.\\
$^{5}$Istituto Nazionale di Fisica Nucleare (INFN),  Sezione di Napoli,\\
Complesso Universitario di Monte Sant'Angelo, Edificio G, Via Cinthia, I-80126, Napoli, Italy.\\
$^{6}$Gran Sasso Science Institute, viale F. Crispi 7, I-67100, L'Aquila, Italy.\\
$^{7}$Laboratory for Theoretical Cosmology,\\
Tomsk State University of Control Systems and Radioelectronics (TUSUR), 634050 Tomsk, Russia.
}

\date{Accepted XXX. Received YYY; in original form ZZZ}

\pubyear{2019}

\begin{document}

\label{firstpage}
\pagerange{\pageref{firstpage}--\pageref{lastpage}}

\maketitle

\begin{abstract}
The assumption of homogeneity and isotropy on large scales is one of the main hypotheses of the standard cosmology. In this paper, we test the hypothesis  of isotropy  from the two-point angular correlation function of 2626 gamma-ray bursts (GRB) of the FERMI GRB catalogue. We show that the uncertainties in the GRB positions induce spurious anisotropic signals in their sky distribution. However, when such uncertainties are taken into account no significant evidence against the large-scale statistical isotropy is found. This result remains valid even for the sky distribution of short-lived GRB, contrarily to previous reports. 

\end{abstract}




\section{Introduction}

One of the foundations of modern cosmology is the so-called Cosmological Principle (CP), which consists in the assumption that the Universe looks homogeneous and isotropic on large scales. Statistical analyses using recent cosmological observations bring evidence that the CP holds true at such scales, as obtained from the Cosmic Microwave Background (CMB) temperature anisotropies~\citep{Ade:2015hxq}, cosmic distances from type Ia Supernovae~\citep{Andrade:2017iam, Deng:2018jrp, Sun:2018cha, Andrade:2018eta, Zhao:2019azy, Soltis:2019ryf}, galaxy number counts~\citep{Gibelyou:2012ri, Yoon:2014daa, Bengaly:2017zlo, Rameez:2017euv}, the sky distribution of galaxy clusters~\citep{Bengaly:2015xkw, Migkas:2017vir}. There is also evidence for a homogeneity scale in the counts of quasars and galaxies~\citep{Scrimgeour:2012wt, Ntelis:2017nrj, Goncalves:2017dzs, Goncalves:2018sxa}. However, some controversial claims have appeared in the literature, such as large-angle features in the CMB~\citep{Schwarz:2015cma} and a large dipole anisotropy in radio source counts~\citep{Singal:2011dy, Rubart:2013tx, Bengaly:2017slg} (see also~\citealt{Dolfi:2019wye}.).

Gamma-ray bursts (GRB) have also been used to test the CP. These events are extremely energetic explosions, whose range lies between $10^{50} - 10^{54}$ erg, which exceeds hundred times the total energy radiated by a supernova. Also if they are not properly {\it standard candles}, they may reveal themselves as possible formidable distance indicators. For a detailed  discussion on the topic, we refer the reader to~\citealt{2013IJMPD..2230028A, 2018PASP..130e1001D, Dainotti:2016qxe}. They are usually classified into short-lived ($T_{90} < 2s$, SGRBs) and long-lived ($T_{90} > 2s$, LGRBs), where $T_{90}$ denotes the duration in which $90\%$ of the burst fluence is accumulated. The LGRBs have been thought to originate from distance star-forming galaxies and associated with collapse of massive stars related to a supernova (SN). However, LGRBs with no clear association with any bright SN have been found. Besides, some LGRBs were observed in a very metal-rich systems, in contradiction to the core-collapse model~\citep{Woosley:2006fn}. Hence, the formation of LGRBs besides the collapse scheme is still a mater of debate. SGRBs are attributed to mergers between neutron star-black hole (NS-BH) or NS-NS. However, \citet{Perna:2016jqh} proposed an alternative scenario that can also lead to a SGRB from mergers of BH-BH. Therefore, SGRBs have been correlated with the local Universe \citep{Tanvir:2005cx, Ghirlanda:2005ev}, while LGRBs typicall consist of higher redshift objects, $ z\sim 9.4$~\citep{2011ApJ...736....7C}.

In the last decades, several authors have shown that the GRBs sky distribution is consistent with statistical isotropy~\citep{hartmann1989angular, hartmann1991fuzzy, Meegan:1992xg, Briggs:1995bk, Tegmark:1995hi}. However, subsequent works suggested otherwise for SGRBs~\citep{Balazs:1998tt, Meszaros:2000ct, Magliocchetti:2003uy, Vavrek:2008ha, Tarnopolski:2015ksu}. Moreover, there are also claims for the  existence of GRB structures of $\sim$ 2000 Mpc at $z \sim 2$~\citep{Horvath:2014wga, Horvath:2015axa, Ruggeri:2016ooh} -- for a different conclusion, see also~\citep{Bernui:2007br, Gibelyou:2012ri, Li:2015yha, Ukwatta:2015rxa, Tarnopolski:2015ksu, Ripa:2017scm, Ripa:2018hak}\footnote{In addition to test the CP, GRB have also been used to probe fundamental physics~\citep{Petitjean:2016ydt}.}.

Given the relevance of the topic, and the current controversies, we revisit in this paper the question of the statistical isotropy in the GRB sky distribution. The analysis performed uses the  two-point angular correlation function (2pACF) of 2627 gamma-ray bursts of the Fermi Gamma-ray Burst Monitor Burst catalogue. We show that, after removing objects with large sky positional errors, there is no evidence of anisotropy signatures for the whole GRB sample, as well as for LGRBs and SGRBs sub-samples. We also applied our data analysis to the Burst And Transient Source Experiment (BATSE) current GRB catalogue of the Compton Gamma Ray Observatory (CGRO) satellite containing 2702 GRBs and similar results are found.

This paper is organised as follows. In Sec. II the observational data set used in the analysis is discussed. The methodology and data analysis performed are presented in Sec. III. Sec. IV presents our main results whereas Sec. V discusses such results and summarises our main conclusions.

\section{The observational data set}

In our analysis we use the Fermi Gamma-ray Burst Monitor Burst Catalogue, termed FERMIGBRST~\citep{Gruber:2014iza,vonKienlin:2014nza,Bhat:2016odd}.
This catalogue is one of the most complete GRB catalogues currently available, comprising 2627 objects detected from July 14th 2008 until August 06th 2019~\footnote{\url{https://heasarc.gsfc.nasa.gov/W3Browse/fermi/fermigbrst.html.}}. Specifically, we make use of the following quantities:

\begin{enumerate}
\item RA: The Right Ascension of the burst, given in J2000 decimal degree.

\item DEC: The Declination of the burst, given in J2000 decimal degree.

\item ErrorRadius: The uncertainty of the object position, in degrees. We term it as $\sigma_r$,  which ranges from zero, when the source localisation was obtained using something other than Fermi GBM (for example, Swift, XMM), up to $ \sim 68^\circ $.

\item $T_{90}$: The duration, in seconds, during which $90\%$ of the burst fluence was accumulated.

\end{enumerate}

For completeness, we also use the BATSE current Gamma-Ray Burst catalogue, which contains 2702 GRBs detected from March 21th 1991 until May 26th 2000\footnote{\url{https://heasarc.gsfc.nasa.gov/W3Browse/all/batsegrb.html.}}. We make use of similar quantities as the FERMIGRBST catalogue.

\begin{figure}


\includegraphics[width=\columnwidth]{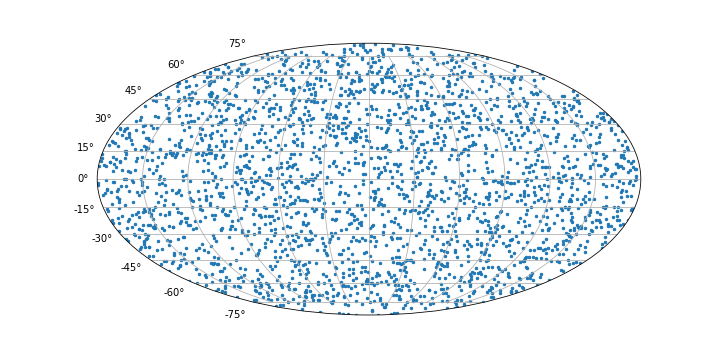}

    \caption{Mollweide projection of the GRB sky positions of the FERMIGBRST catalogue.}
    \label{fig:fermigbrst}
    
    \includegraphics[width=\columnwidth]{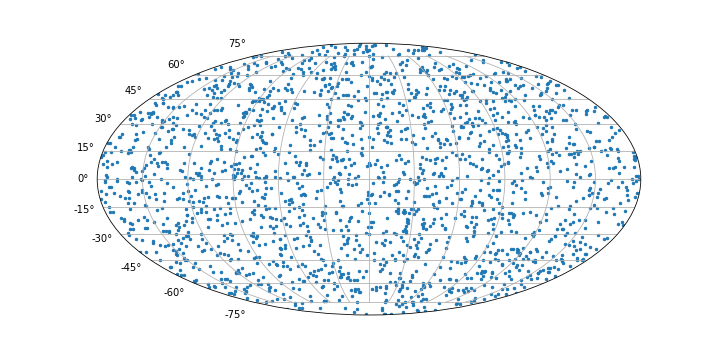}
    \caption{ The same as Fig.~\ref{fig:fermigbrst}, but for the CGRO/BATSE catalogue}
    \label{fig:BATSE}
    
\end{figure}

The FERMIGRBST and CGRO/BATSE sky distribution of the selected GRBs are displayed in the Fig.~\ref{fig:fermigbrst} and Fig.~\ref{fig:BATSE}, respectively. Note that we removed one object in the former and 665 objects in the latter, because they do not contain information on the $T_{90}$ parameter. We test the statistical isotropy for three cases, namely the whole GRB sample, (allGRB), as well as the LGRB and SGRB sub-samples.

\section{Data analysis}

\subsection{Two-point angular correlation function}
\label{sec:2pACF} 

\begin{figure*}
\includegraphics[scale=0.75 ]{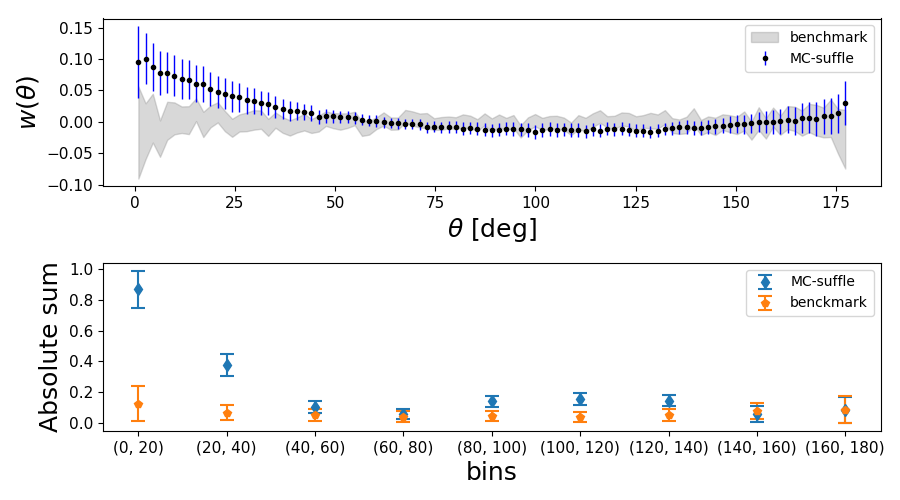}

\caption{\textit{ Top}: Comparison between the 2pACF of the MC-shuffle realisations and the benchmark ones. Blue dots represent the former, while shaded regions  correspond to the latter within the standard deviation bounds. \textit{Bottom}: The absolute sum test.  These results were obtained assuming $20^{\circ}$ size bin, and $\sigma_{\rm r} \leq 20^\circ$.
}
\label{fig:mc_shuffle1}
\end{figure*}
\begin{figure*}
\includegraphics[scale=0.75]{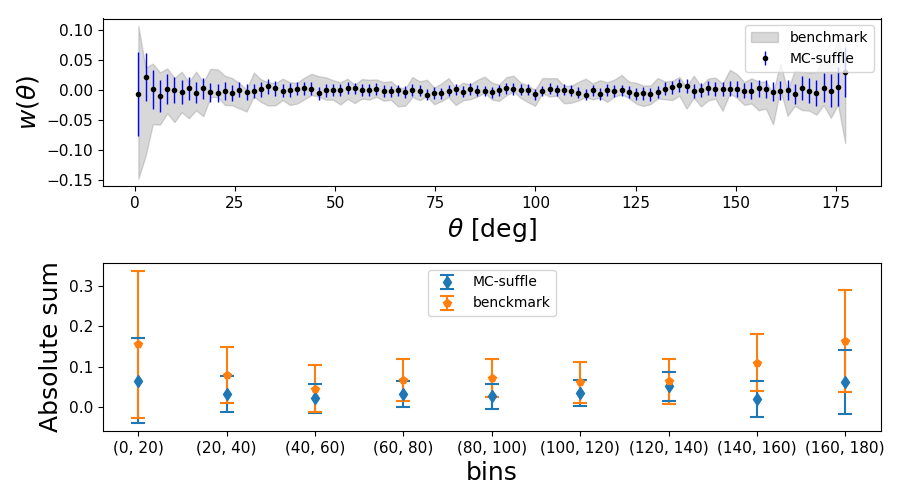}

\caption{The same as Fig.~\ref{fig:mc_shuffle1}, but valid for $\sigma_{\rm r} \leq 6^\circ$.}
\label{fig:mc_shuffle2}
\end{figure*}

In order to summarise the distribution of data points in the sky, one can report the mean number of points at a given scale. However, in the presence of clustering, the mean may be an insufficient descriptor, as this measure is insensitive to it. The 2pACF is a statistic capable of characterising the clustering of objects in the sky, which we denote by $\omega(\theta)$. Essentially, it provides the probability to find a pair of object encompassed in a given solid angle with respect to an isotropic distribution~\citep{1980lssu.book.....P}. As a result, the 2pACF is null in the absence of clustering, and it is statistically similar for all angular scales if the underlying distribution is statistically isotropic.

We can determine the 2pACF from the data using the well-known Landy-Szalay estimator~\citep{Landy:1993yu},

\begin{equation}
    \omega(\theta) = \frac{\langle DD(\theta) \rangle - 2  \langle DR(\theta)\rangle +  \langle RR(\theta) \rangle}{\langle RR(\theta) \rangle}\,,
	\label{eq:2pACF-LS}
\end{equation}
where the brackets denote the normalised number of all GRB pairs in the real data ($DD(\theta)$), in the auxiliary random isotropic catalogue ($RR(\theta)$), and between the data and the random catalogue ($DR(\theta)$). The counts of pairs for each angular scale is carried out through the range $(\theta - d\theta/2, \theta + d\theta/2)$, where $d\theta$ is the bin width. We choose evenly spaced samples in our analysis, so that $\omega(\theta)$ is calculated over the interval $(0^\circ, 180^\circ)$ with $d\theta = 1.8^\circ$. The random isotropic catalogues are generated following a uniform distribution on a sphere, so that

\begin{eqnarray}
    \mathrm{RA} &=& 0^\circ + (360^\circ - 0^\circ) \ U[0, 1), \\
    \mathrm{DEC} &=& \arcsin\left( -90^\circ + \ (180^\circ) \ U[0, 1) \, \right) \,.
	\label{eq:uniform_sky}
\end{eqnarray}

We  estimate our uncertainties of the 2pACF via bootstrap method~\citep{Tarnopolski:2015ksu}.  By means of 100 re-samplings between the data and the random catalogue, we obtain a bootstrap sampling distribution of the 2pACF whose uncertainty is the standard deviation of this sampling distribution.      
We also compute the 2pACF for isotropic random sample through Monte Carlo (MC) method, and perform the bootstrap analysis on it as well. We denote the mean and the standard deviation of this distribution as the {\it benchmark}~\citep{Tarnopolski:2015ksu}, since it provides the 2pACF limits that a finite isotropic sample must satisfy. Any significant deviation from this benchmark would hint at departures from statistical isotropy.

\subsection{Absolute sum test}\label{subsec:somatorio}

In order to further investigate the readability of the anisotropic signal, we devised what we hereafter call ``Absolute sum test". We defined it as the summation of the absolute 2pACF values within a specific angular scale, i.e.,

\begin{eqnarray}
\mathrm{Absolute \;\; sum} \equiv \sum_i |\omega(\theta)|_i \; \forall i \in \Delta\theta,
	\label{eq:AbsoluteSumTest}
\end{eqnarray}

\noindent where $\Delta\theta$ is the angular range at which we split the $\omega(\theta)$, so that we perform this summation in a tomographic fashion within this range, i.e., $i_1 = (0^{\circ}, 20^{\circ}), i_2 = (20^{\circ},40^{\circ}), \dots, i_9 = (160^{\circ}, 180^{\circ})$. We choose $\Delta\theta=20^{\circ}$ in order to have at least 10 values of $\omega(\theta)$ within this interval (since $d\theta=1.8^{\circ}$) and thus a robust statistics for each $i$-bin. 
Hence we can compare the 2pACF at different angular scales for two distributions, e.g. the real data versus the benchmark. Again, any deviation beside the error bars (as obtained from standard uncertainty propagation) suggests potential deviations from statistical isotropy, after all, the 2pACF values should agree at different angular if this assumption holds.

\subsection{Statistical significance estimate}\label{subsec:statsig}

We assess the statistical significance of our analysis by means of non-parametric tests between two different samples as follows:
\begin{enumerate}

\item Kolmogorov-Smirnov (KS): It consists of a distribution-free test which compares the empirical cumulative distribution function (ECDF) of two samples~\citep{astroMLText}. This test relies on a metric which measures the maximum distance of the two ECDF $F_m(x)$ and $G_n(x)$,  

\begin{eqnarray}
    D = \max\left |F_{\rm m}(x) - G_{\rm m}(x)\right | \,.
	\label{eq:K-S}
\end{eqnarray}

\item Anderson-Darling (AD): This test is based on the following statistics~\citep{scholz1987k}:

\begin{eqnarray}
    A^{2}_{\rm mn} = \frac{mn}{N} \int_{-\infty}^{\infty} \frac{\left\lbrace F_{\rm m}(x) - G_{\rm n}(x)\right\rbrace^2}{H_{\rm N}(x) \left\lbrace 1 - H_{\rm N}(x)\right\rbrace} dH_{\rm N}(x) \,.
	\label{eq:A-D}
\end{eqnarray}

\noindent $F_{\rm m}$ and $G_{\rm n}$ are ECDF for two independent samples that may have different number of points, namely $n$ and $m$, respectively. $H_{\rm N}(x) = \left\lbrace m F_{\rm m} + n G_{\rm n}\right\rbrace /N $ is the ECDF of the pooled sample, where $N = m+n$.

\end{enumerate}

Our null hypothesis is that the two sample are drawn from the \textit{same} distribution. KS test is sensitive to the location, the scale and the shape of the distribution, while AD test is only sensitive to the shape. Moreover, AD test is more sensitive to the tails differences than KS test, which in turn is more sensitive to the differences near the centre of the distribution. In this sense, these two non-parametric tests are complementary.

We choose $\alpha = 0.05$ as the significance level in which we reject the null hypothesis. Hence, a p-value {\it lower} than $\alpha$ when we compare the real data with the benchmark, for example, would denote that the samples are not drawn from the same distribution - and thus the data is not statistically isotropic. We used the routines \sc{ks\_2samp} \rm{and} \sc{anderson\_ksamp} \rm of the \sc{SciPy} \sc{Python} \rm library to compute the KS and AD statistics, as well as the p-values\footnote{Jones E, Oliphant E, Peterson P, et al. SciPy: Open Source Scientific Tools for Python, 2001-, http://www.scipy.org/ [Online; accessed 2018-11-13].}.

\begin{figure*}
\includegraphics[scale=.5]{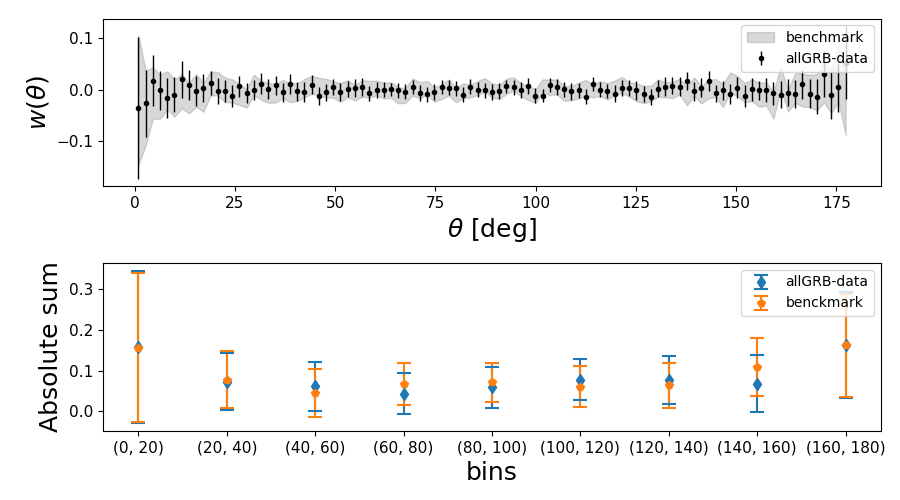}

\caption{\textit{Top}: The 2pACF obtained from a synthetic isotropic sample (benchmark; shaded regions), and from the GRBs FERMIGBRST catalogue with $\sigma_{\rm r} \leq 6^{\circ}$ (black dots). \textit{Bottom}: The absolute sum of the 2pACF per bin, again with bin size. The diamond markers correspond to the real data, while pentagon markers represent the isotropic realisations.  These results are valid for the FERMIGBRST allGRB sample.}

\label{fig:data_vs_benck1}
\end{figure*}
\begin{figure*}
\includegraphics[scale=.5]{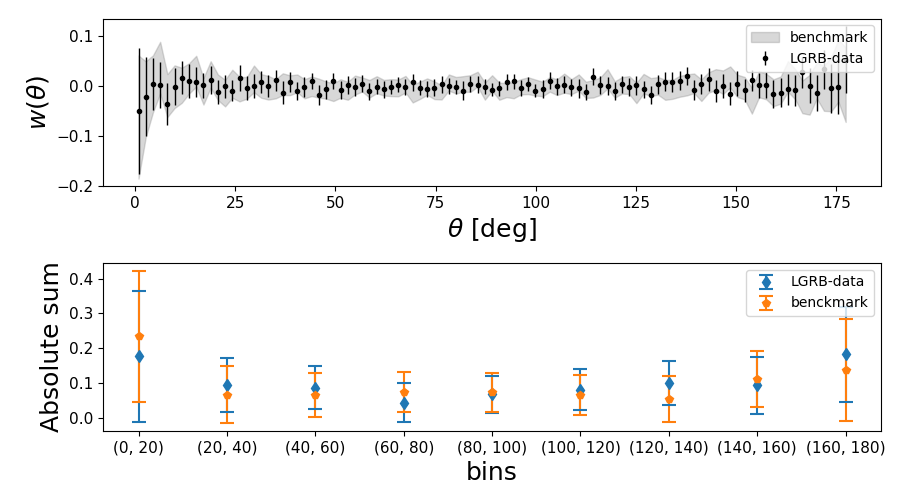}

\caption{ The same as Fig.~\ref{fig:data_vs_benck1}, but for FERMIGBRST LGRB sample.} 
\label{fig:data_vs_benck2}
\end{figure*}
\begin{figure*}
\includegraphics[scale=.5]{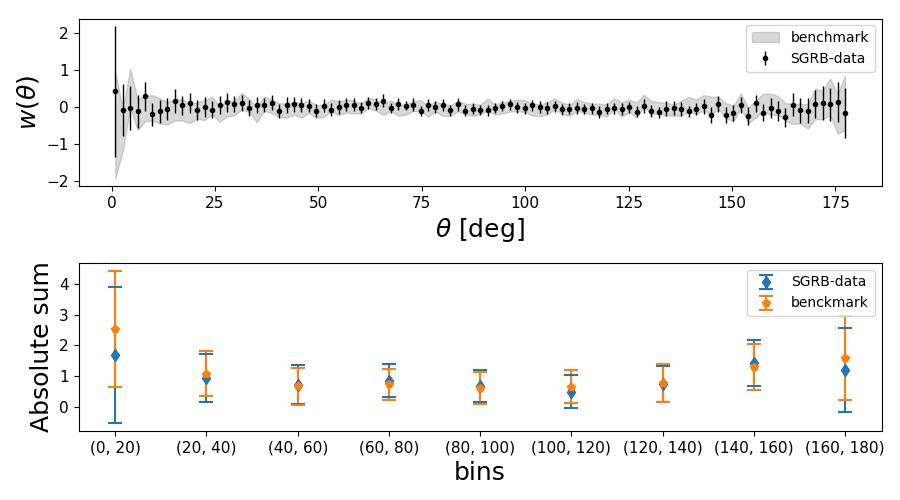}
\caption{ The same as Fig.~\ref{fig:data_vs_benck1}, but for FERMIGBRST SGRB sample.} 
\label{fig:data_vs_benck3}
\end{figure*}

\begin{figure*}
\includegraphics[scale=.5]{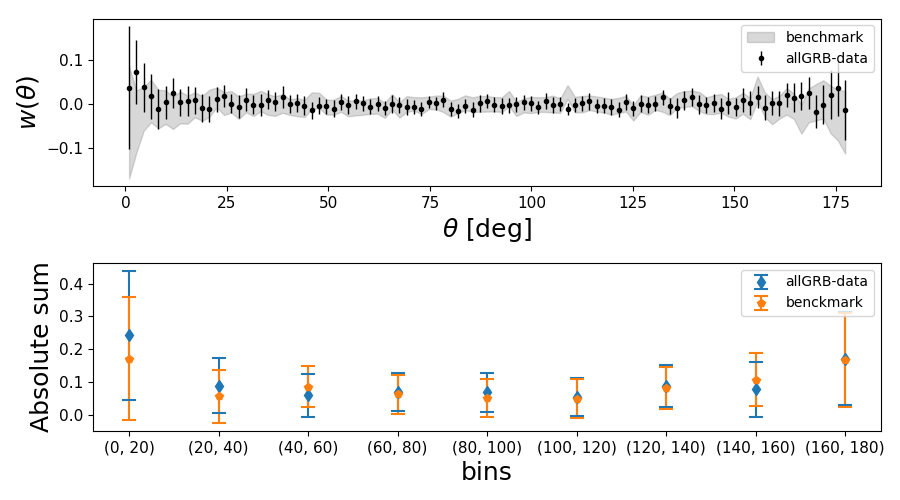}
\caption{The same as Fig.~\ref{fig:data_vs_benck1}, but for the CGRO/BATSE allGRB sample.} 
\label{fig:data_vs_benck_BATSE1}
\end{figure*}
\begin{figure*}
\includegraphics[scale=.5]{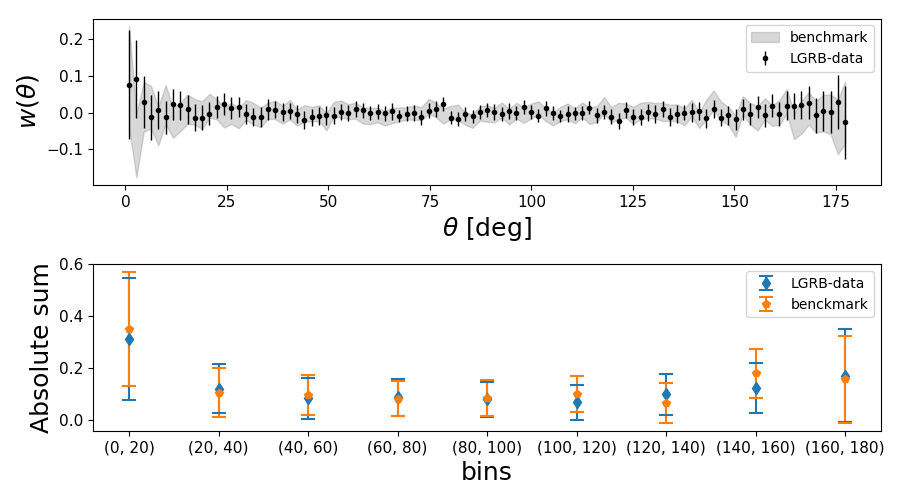}
\caption{The same as Fig.~\ref{fig:data_vs_benck_BATSE1}, but for CGRO/BATSE LGRBs} 
\label{fig:data_vs_benck_BATSE2}
\end{figure*}
\begin{figure*}
\includegraphics[scale=.5]{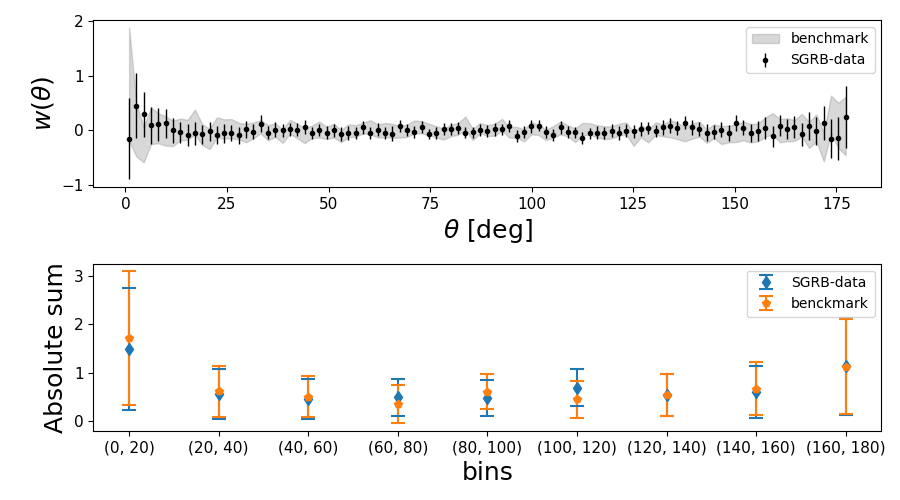}
\caption{The same as Fig.~\ref{fig:data_vs_benck_BATSE1}, but for CGRO/BATSE SGRBs} 
\label{fig:data_vs_benck_BATSE3}
\end{figure*}

\subsection{GRB positional uncertainties}

Previous works showed that the GRB positional uncertainties might affect the measurement of the 2pACF and angular power spectrum of their celestial distribution~\citep{hartmann1991fuzzy,Tegmark:1995hi}. Here, in order to investigate the impact of such uncertainties on our analysis, we produce 1000 Monte Carlo (MC) realisations with the following prescription:

\begin{eqnarray}
    \mathrm{RA_{new}} &=& \mathcal{N}\left(\mathrm{RA}, \sigma_{\rm r}^2\right); \\
    \mathrm{DEC_{new}} &=& \mathcal{N}\left(\mathrm{DEC}, \sigma_{\rm r}^2 \right) \,.
	\label{eq:MC-shuffle}
\end{eqnarray}

Due to the shuffling of GRB celestial positions, we henceforth call this test {\it MC-shuffle}. We compute the 2pACF for each of these realisations, then we take their mean and standard deviation, and compare them with the benchmark. We repeat this procedure for different positional uncertainty cutoffs, which we denote by $\sigma_{\rm r}$. If we find significant discrepancy between these MC-shuffle and benchmark realisations for less restrictive cutoffs, this  will hint at spurious anisotropies arising due to such uncertainties. 

\section{Results}

We depict the impact of GRB positional errors on the 2pACFs in Fig.~\ref{fig:mc_shuffle1} and Fig.~\ref{fig:mc_shuffle2}. By comparing the MC-shuffle and the benchmark with an upper limit of $\sigma_{\rm r}=20^\circ$ and $\sigma_{\rm r}=6^\circ$, we can clearly see that the  latter is in better agreement with the benchmark, especially on smaller angular scales. This result shows that large positional errors indeed introduce spurious anisotropic signatures, as previously reported in~\cite{hartmann1991fuzzy,Tegmark:1995hi}, and therefore we should impose an upper $\sigma_{\rm r}$ cutoff in our working sample. The GRB sub-sample with $\sigma_{\rm r} \leq 6^\circ$ will be hereafter taken as our real data sample, since it alleviates this feature, and we do not lose a large fractional of GRB, as we still retain 1760 GRBs in total, i.e., 1590 LGRBs and 170 SGRBs. 

We compare the 2pACF between the real data and the benchmark in the upper panels of Fig.~\ref{fig:data_vs_benck1}, while the bottom ones display the results from the absolute sum test as described in  Section~\ref{subsec:somatorio}. The results obtained for LGRBs and SGRBs samples are displayed, respectively, in Fig.~\ref{fig:data_vs_benck2} and~\ref{fig:data_vs_benck3}. Once more the shaded regions in the upper panels provide the allowed region which an intrinsic isotropic sample can vary due to randomness. From theses plots, we can conclude that both the 2pACF and the absolute sum show a good agreement between the allGRBs, LGRBs and SGRBs and the benchmark, and therefore no evidence against statistical isotropy. We verified that our results are unchanged for sub-samples of GRBs with fluence above a given value as $5 \times 10^{-6} \; {\rm erg \, cm}^{-2} $ or $10^{-5} \; {\rm erg \, cm}^{-2}$
 
In addition, we show the results obtained from the non-parametric tests in Table~\ref{tab:nonparametrictest}. We obtain that we cannot reject the null hypothesis at significance level of $\alpha = 0.05$ for all data samples, including SGRBs, although they yield the lowest p-value among all. We then confirm that the sky distribution of FERMIGBRST catalogue of GRBs is statistically isotropic, as expected from the CP. 

For completeness, we also report the same analysis for the CGRO/BATSE GRB catalogue in Figs.~\ref{fig:data_vs_benck_BATSE1},~\ref{fig:data_vs_benck_BATSE2} and~\ref{fig:data_vs_benck_BATSE3} for the allGRBs, LGRBs and SGRBs, respectively. We follow \citet{chen1998two} who found that sky-exposure effects are small and should not affect our results. Here again, we see an agreement with the benchmark, which means no departure from statistical isotropy. The non-parametric tests for the CGRO/BATSE data set are displayed in Table~\ref{tab:nonparametrictest_BATSE}. Although this set shows consistency with statistical isotropy, the allGRB sample shows departure from isotorpy. We discuss this latter result in the next section.

\begin{table}
	\centering
	\caption{Results using both the Kolmogorov-Smirnov statistic D and the Anderson-Darling statistic AD  for the FERMIGBRST catalogue.}
    
	\label{tab:nonparametrictest}
	\begin{tabular}{lccr} 
		\hline
		& allGRBs & LGRBs & SGRBs\\
		\hline
		D & 0.11 & 0.11 & 0.14\\
		p-value & 0.55 & 0.55 & 0.26\\
		\hline
		AD & -0.36\tablefootnote{Negative values of AD statistic means that the p-value was obtained as an upper limits and the true value might be larger than 0.25} & -0.88 & 0.37 \\
    	p-value & $\geq$0.25 & $\geq$ 0.25 & 0.24 \\

		\hline
	\end{tabular}
\end{table}

\begin{table}
	\centering
	\caption{Results using both the Kolmogorov-Smirnov statistic D and the Anderson-Darling statistic AD  for the CGRO/BATSE catalogue.}
    
	\label{tab:nonparametrictest_BATSE}
	\begin{tabular}{lccr} 
		\hline
		& allGRBs & LGRBs & SGRBs\\
		\hline
		D & 0.21 & 0.11 & 0.12\\
		p-value & 0.02 & 0.55 & 0.44\\
		\hline
		AD & 4.3 & 0.17 & -0.24 \\
    	p-value & 0.006 & $\geq$ 0.25 & $\geq$ 0.25 \\

		\hline
	\end{tabular}
\end{table}

\section{Discussion and conclusion}

One of the major concepts of modern cosmology is the assumption of the statistical homogeneity and isotropy on large scales.  Together with the Einstein's field equations, they are at the basis of what we know as modern cosmology.  

In this paper, we probed the statistical isotropic hypothesis of the CP by means of the 2pACF of the GRB sky distribution. To perform our analysis, we compared the 2pACF of the Fermi GRB catalogue with the 2pACF of the isotropic synthetic sample.  We also investigated how the uncertainty in the GRBs position might affect ours conclusion by drawing 1000 MC simulations with new GRB positions inside the radius of the observational positional uncertainty. 

We found that large positional uncertainties lead to spurious anisotropy detection, as shown in Fig.~\ref{fig:mc_shuffle1}. For this reason, we perform cuts on the position uncertainty, choosing $\sigma_{\rm r}=6^{\circ}$ as an optimal upper cut in which we can avoid spurious anisotropy without losing too many sources. Then, we split the data in three samples: all GRBs, LGRBs and SGRBs, containing 1760, 1590 and 170 each, respectively, after this cutoff. Fig.~\ref{fig:data_vs_benck1},~\ref{fig:data_vs_benck2}, and~\ref{fig:data_vs_benck3} show the results for each sample, respectively.  We found a good agreement between all these data and the statistical isotropy hypothesis, since the 2pACF and the absolute sum test agrees with the benchmark simulations. This was also confirmed by the KS and AD tests between the real data and the benchmark, whose p-values are {shown} in Table~\ref{tab:nonparametrictest}. None of these p-values were smaller than $0.05$, meaning we cannot reject the null hypothesis at this significance level. We remark that in the case of SGRB, despite a lower p-value, one still cannot reject the null hypothesis. 

We performed the same analysis for the CGRO/BATSE GRB catalogue, whose results are shown in Fig.~\ref{fig:data_vs_benck_BATSE1},~\ref{fig:data_vs_benck_BATSE2} and~\ref{fig:data_vs_benck_BATSE3} for allGRB, LGRB and SGRB samples, respectively, as well as in Table~\ref{tab:nonparametrictest_BATSE}. We conclude that statistical isotropy holds in this catalogue as well, although that the allGRB sample exhibits a stronger departure from this assumption. We credit this anisotropy to the impact of the sky-exposure function. Although \citet{chen1998two} showed that the effect on the 2pACF is small using a simple $\chi^2$ analysis, we argue that our analysis is more sensitive, and thus we could capture the impact of the sky-exposure function in the 2pACF as other anisotropy tests e.g. dipole and quadrupole moments~\citep{1996ApJS..106...65M}. This result shows that our method is robust enough to capture even small anisotropy effects.

We conclude that there is no significant evidence for isotropy departure in the currently available GRB catalogues - even in the the SGRB sub-samples. This result is in good agreement with~\cite{Tarnopolski:2015ksu,Ukwatta:2015rxa,Ripa:2017scm,Ripa:2018hak},  although we used an updated sample of GRB, and a different estimator. Therefore, we confirm the validity of statistical isotropy of the GRB distribution across the sky, which should be definitely underpinned in light of forthcoming Gamma-Ray surveys like e.g. THESEUS~\citep{Amati:2017npy}.
Specifically, this space mission is aimed to exploit GRBs in view of investigating the early Universe and then providing a substantial advance in  time-domain astrophysics. Due to the wide range of redshift that this mission can reach, it can also be extremely interesting for multi-messenger astrophysics as well.


\section*{Acknowledgements}

U.A. acknowledges financial support from CAPES. C.A.P.B. acknowledges financial support from the South African SKA Project. J.S.A. acknowledges support from CNPq (grant Nos. 310790/2014-0 and 400471/2014-0) and FAPERJ (grant No. E-26/203.024/2017). S.C. acknowledges support from INFN ({\it Iniziativa Specifica} QGSKY) and CANTATA COST action CA15117.



\bibliographystyle{mnras}
\bibliography{refs} 







\bsp	
\label{lastpage}
\end{document}